\documentclass{DISproc}

\def\kt{\ensuremath{k_t}}

\newcommand{\asb}{{\bar \alpha}_\mathrm{s}}
\newcommand{\Pmax}{p}
\def\CASCADE{{\sc Cascade}}
\newcommand{\herafitter}{Aaron:2009aa,Aaron:2009kv,herafitter}

\begin{document}
\title{Determination of transverse momentum dependent gluon density from
HERA structure function measurements }

\author{{\slshape Hannes Jung$^{1,2}$, Francesco Hautmann$^3$}\\[1ex]
$^1$DESY, Notkestra{\ss}e 85, 22607 Hamburg, Germany\\
$^2$CERN, 1211 Gen\`eve 23, Switzerland\\
$^3$Theoretical Physics Department, 
University of Oxford,    Oxford OX1 3NP, GB}

\contribID{xy}

\doi  

\maketitle

\begin{abstract}
  The transverse momentum dependent gluon density obtained with CCFM evolution is determined from a fit to the latest combined HERA structure function measurements. 
\end{abstract}

\section{Introduction}
The combined measurements of the structure function at HERA \cite{Aaron:2009aa} allow the determination of parton distribution functions to be carried out to high precision. While these data have been used to determine the collinear parton densities,  the transverse momentum distributions (TMD) or unintegrated gluon distributions were only based on older and much less precise measurements  \cite{Jung:2002wn,Hansson:2003xz}.

In high energy factorization \cite{Catani:1990eg} the cross section is written as a convolution of the partonic cross section $\hat{\sigma}(É \kt)$ which depends on the transverse momentum $\kt$ of the incoming parton with the $\kt$-dependent parton density function ${\cal \tilde A}\left(x,\kt,\Pmax\right)$:
\begin{equation}
 \sigma  = \int 
\frac{dz}{z} d^2k_t \hat{\sigma}(\frac{x}{z},k_t)  {\cal \tilde A}\left(x,\kt,\Pmax\right)\label{kt-factorisation}
\end{equation}
\begin{wrapfigure}{r}{0.28\textwidth}
\vskip -0.7cm
\centering \includegraphics[width=0.15\textwidth]{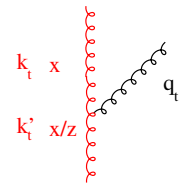}
\vskip -0.3cm
\caption{Gluon branching}
  \label{Fig:branching}
\end{wrapfigure}
where $\Pmax$ is the factorization scale. The evolution of ${\cal \tilde A}\left(x,\kt,\Pmax\right)$ can proceed via the BFKL, DGLAP or via the CCFM evolution equations. Here, an extension of the CCFM evolution is applied (to be also used in the parton shower Monte Carlo event generator CASCADE \cite{Jung:2010si}) which includes the use of  two loop $\alpha_s$ as well as applying a consistency constraint \cite{Kwiecinski:1996td,Ciafaloni:1987ur,Andersson:1995ju} in  the  $g\to gg$ splitting function \cite{Andersson:2002cf}:
\begin{equation}
P_{gg}(z,\Pmax,\kt ) = \asb \left(\kt^2 \right) 
\left( \frac{(1-z)}{z} + \frac{z(1-z)}{2}\right) \Delta_{ns} 
  + \asb (\Pmax^2) \left(\frac{z}{1-z} + 
\frac{z(1-z)}{2}\right) ,
\label{fullsplitt}
\end{equation}
with $ \Delta_{ns} $ being the non-Sudakov form factor.
The consistency constraint is given by \cite{Kwiecinski:1996td} (see Fig.~\ref{Fig:branching}):
\begin{equation}
q_t^2   <   \frac{(1-z) \kt^2}{ z}   \label{consistency-constraint}
\end{equation}

\section{Evolution}
Since the CCFM evolution cannot be easily written in an analytic closed form, a Monte Carlo method, based on \cite{Marchesini:1990zy,Marchesini:1992jw}, is used. However, the Monte Carlo solution is time consuming, and cannot be used in a straightforward way in a fit program. For a realistic solution, first a kernel $ {\cal \tilde A}\left(x'',\kt,\Pmax\right) $ is determined from the MC solution of the CCFM evolution equation, and then is folded with the non-perturbative starting distribution ${\cal A}_0 (x)$:
\begin{eqnarray}
x {\cal A}(x,\kt,\Pmax) &= &x\int dx' \int dx'' {\cal A}_0 (x) {\cal \tilde A}\left(x'',\kt,\Pmax\right)  \delta(x' \cdot x'' - x) \\
&= &\int dx' \int dx'' {\cal A}_0 (x) {\cal \tilde A}\left(x'',\kt,\Pmax\right) \frac{x}{x'} \delta(x'' - \frac{x}{x'}) \\
& = & \int dx' {{\cal A}_0 (x') }  
\cdot \frac{x}{x'}{ {\cal \tilde A}\left(\frac{x}{x'},\kt,\Pmax\right) } 
\end{eqnarray}
The kernel  ${\cal \tilde A}$ includes all the dynamics of the evolution, Sudakov form factors and splitting functions and is determined in a grid of $50\otimes50\otimes50$ bins in $x,\kt,\Pmax$.  

The calculation of the cross section according to eq.(\ref{kt-factorisation}) involves a multidimensional Monte Carlo integration which is time consuming and suffers from numerical fluctuations, and cannot be used directly in a fit procedure involving the calculation of numerical derivates in the search for the minimum. Instead the following procedure is applied:
\begin{eqnarray}
\sigma_r(x,Q^2) & = & \int_x^1 d x_g {\cal A}(x_g,\kt,\Pmax) \hat{ \sigma}(x,x_g,Q^2) \\
 & = & \int d x_g\; dx'\; dx'' {\cal A}_0 (x') {\cal \tilde A}(x'',\kt,\Pmax)\cdot \hat{ \sigma}(x,x_g,Q^2) \cdot \delta(x' \,x'' -x_g) \\
 & = & \int dx'\; dx'' {\cal A}_0 (x') \cdot {\cal \tilde A}(x'',\kt,\Pmax) \cdot \hat{ \sigma}(x,x'\,x'',Q^2) \\
 & = & \int_x^1 dx' {\cal A}_0 (x') \cdot \int_{x/x'}^1 dx''  {\cal \tilde A}(x'',\kt,\Pmax) \cdot \hat{ \sigma}(x,x'\,x'',Q^2) \\
  & = & \int_x^1 dx' {\cal A}_0 (x') \cdot \tilde{ \sigma}(x/x',Q^2) \label{final-convolution}
 \end{eqnarray}
Here, first $ \tilde{ \sigma}(x',Q^2)$ is calculated numerically with a Monte Carlo integration on a grid in $x$ for the values of $Q^2$ used in the fit. Then the last step (i.e. eq.(\ref{final-convolution})) is performed with a fast numerical gauss integration, which can be used in standard fit procedures.

The fit to the HERA structure function measurements is performed applying the \verb+herafitter+  program~\cite{\herafitter} to determine the parameters of the starting distribution ${\cal A}_0$  at the starting scale $Q_0$:
\begin{eqnarray}
x{\cal A}_0(x,\kt) &=& N x^{-B_g} \cdot (1 -x)^{C_g}\left( 1 -D_g x\right) 
\label{a0}
\end{eqnarray}
\section{Fit to HERA structure function}
The parameters $N,B_g,C_g,D_g$ in eq.(\ref{a0}) are determined from a fit to the combined structure function measurement  \cite{Aaron:2009aa} in the range $x<0.01$ and $Q^2>5$~GeV. In addition to the gluon induced process $\gamma^* g^* \to q\bar{q}$ the contribution from valence quarks is included via $\gamma^* q \to q$ using a CCFM evolution of valence quarks as described in \cite{Deak:2010gk}. The results presented here are obtained with the \verb+herafitter+ package, treating the correlated systematic uncertainties separately from the uncorrelated statistical and systematic uncertainties.
To obtain a reasonable fit to the structure function data, the starting scale $Q_0$ as well as $\Lambda_{qcd}$ has been varied. An acceptable $\chi^2/ndf $ could only be achieved when applying the consistency constraint eq.(\ref{consistency-constraint}): without consistency constraint the best  $\chi^2/ndf \sim 14 - 28$, depending on which form of the splitting function is used. With consistency constraint and 
the splitting function eq.(\ref{fullsplitt}) the best fit gives $\chi^2/ndf \sim 1.5$ for $Q_0=1.8$~GeV and $\Lambda_{qcd} = 0.17$~GeV at $n_f=4$ flavours. 
It has been checked, that the  $\chi^2/ndf$ does not change significantly when using 3 instead of 4 parameters for the initial starting distribution ${\cal A}_0$.

\begin{figure}[htb]
  \includegraphics[width=0.45\textwidth,bb= 00 120 500 700]{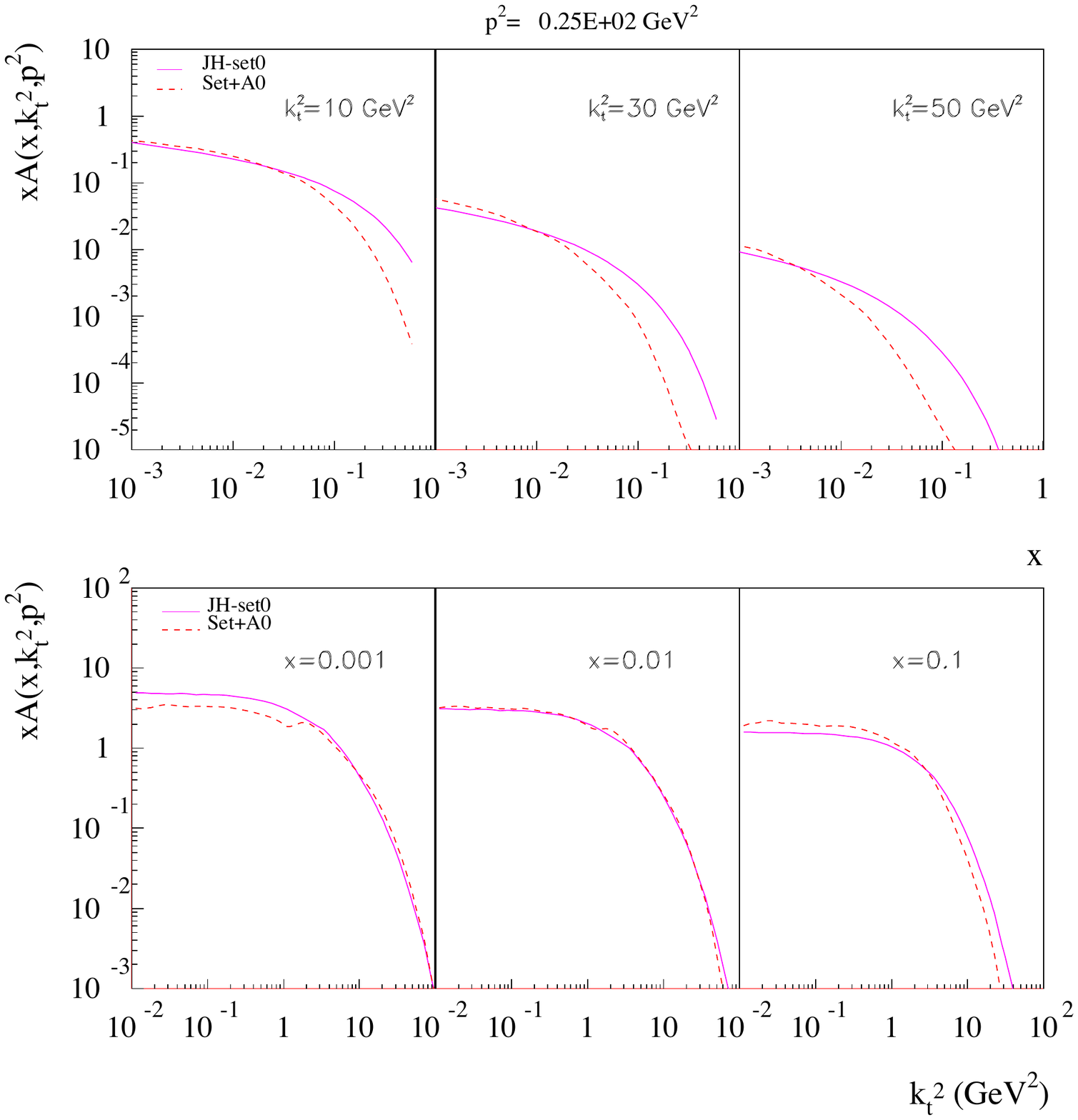}
  \hskip 0.6cm
  \includegraphics[width=0.45\textwidth,bb= 00 120 500 700]{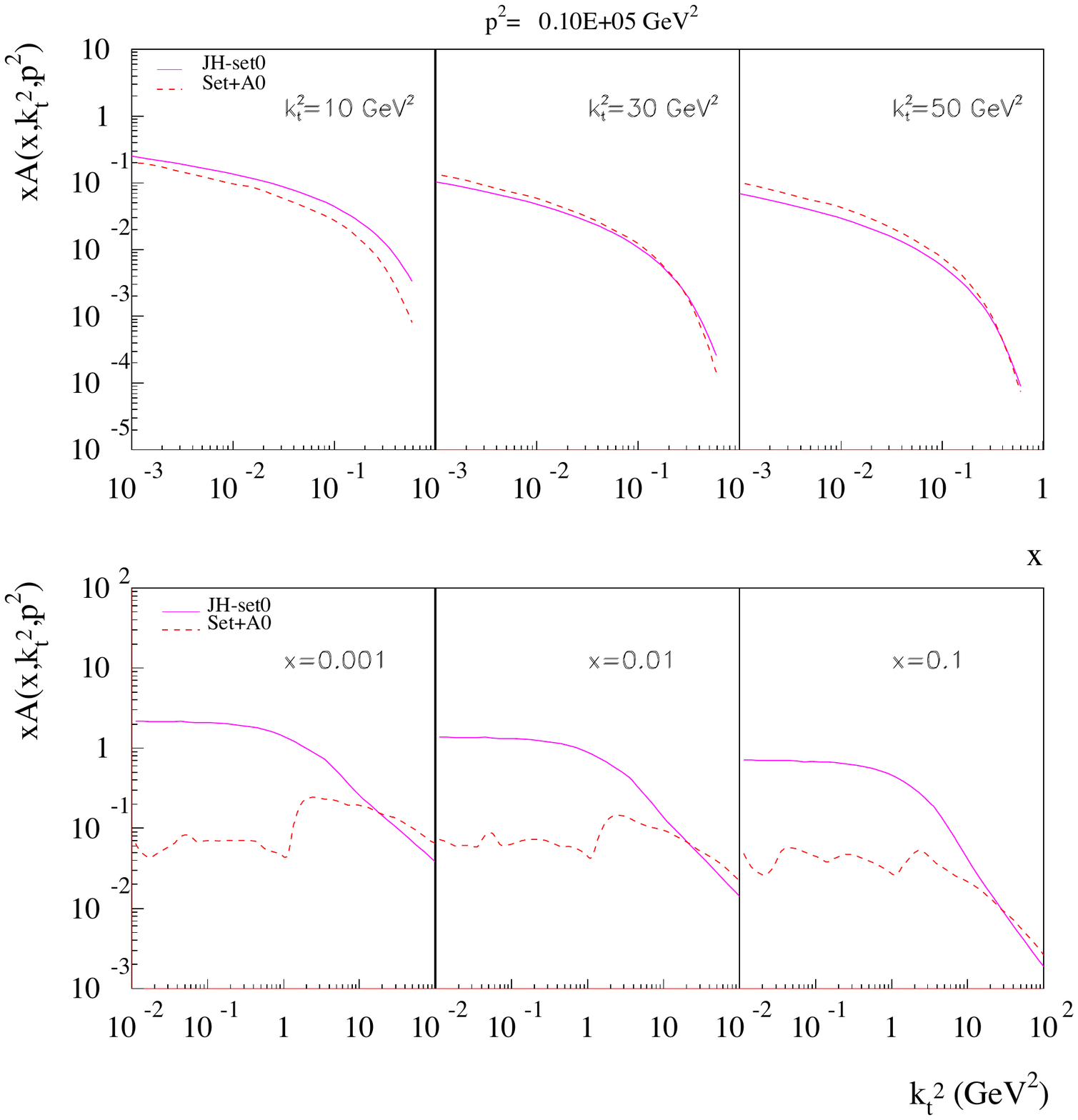}
  \caption{Unintegrated gluon density {\bf JH-set0}  for $\Pmax^2= 25$~GeV$^2$ (left) and $\Pmax^2= 10^5$~GeV$^2$ (right)  as a function of $x$ for different values of $\kt^2$ and as a function of $\kt^2$ for different values of $x$ compared to  {\bf set A0} \protect\cite{Jung:2004gs}}
  \label{Fig:updf}
\end{figure}

In fig.\ref{Fig:updf} the resulting unintegrated gluon density {\bf JH-set0} is shown for 2 values of $\Pmax^2$  compared to  {\bf set A0} \cite{Jung:2004gs}.

\begin{figure}[htb]
  \includegraphics[width=0.45\textwidth,bb= 100 220 500 700]{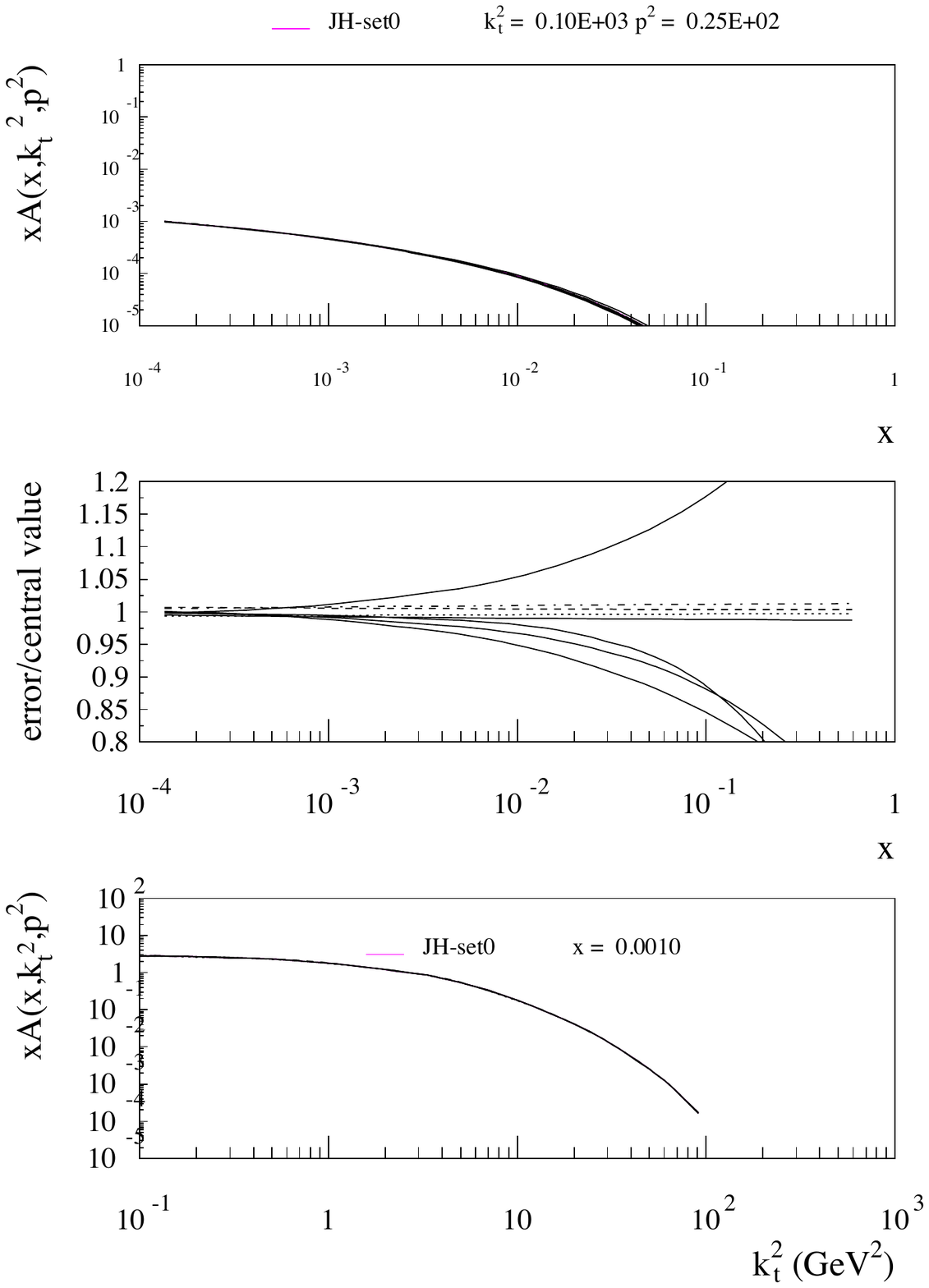}
  \includegraphics[width=0.35\textwidth,bb=00 0 500 700]{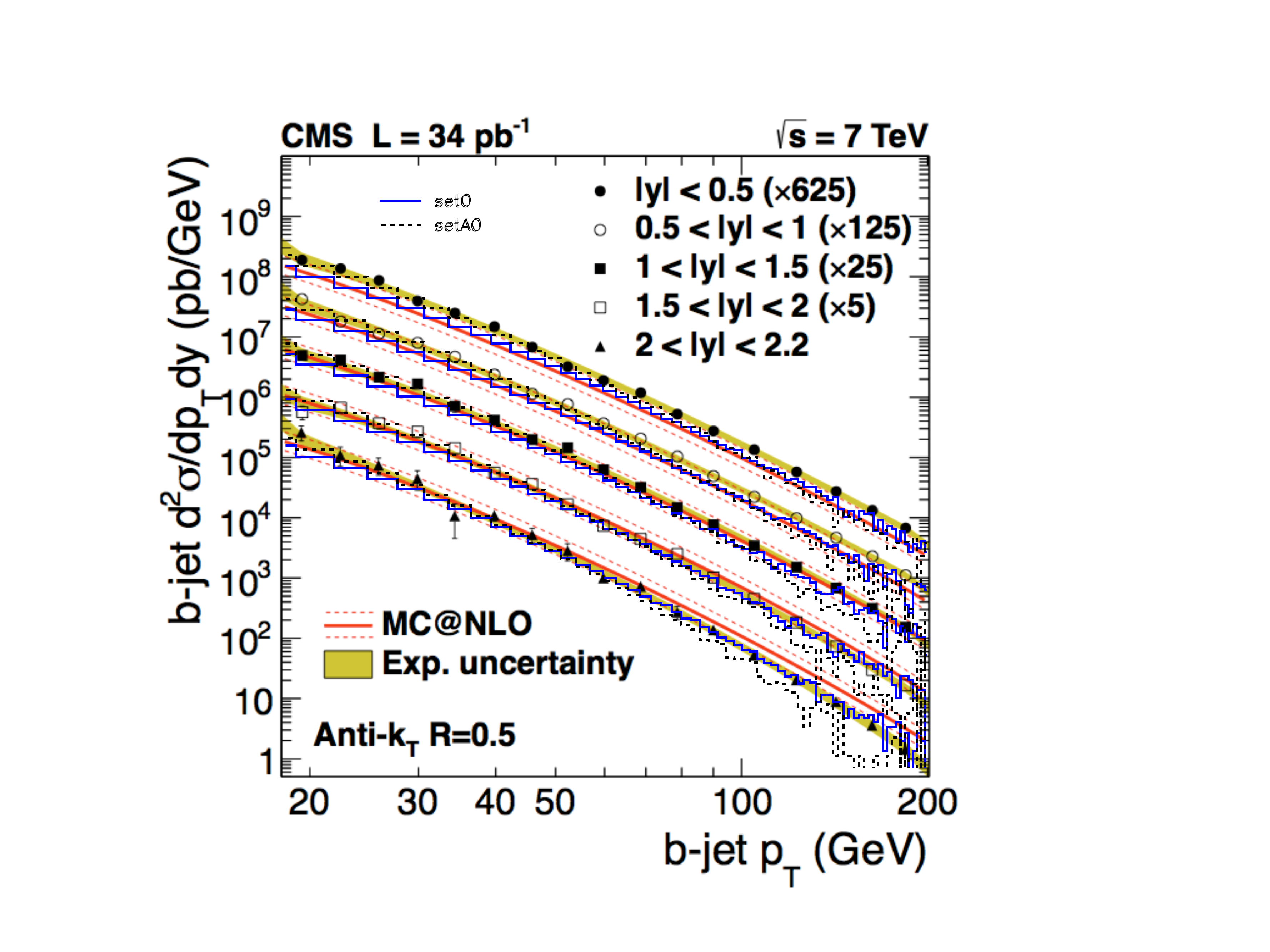}
  \caption{(left): Uncertainties of the uPDF at $\Pmax^2= 25$~GeV$^2$. 
  (right): Cross section of $b$-jet production as a function of $p_t$ for different bins in $y$ as measured by CMS~\protect\cite{Chatrchyan:2012dk} compared to predictions from \CASCADE\ \protect\cite{Jung:2010si} using the unintegrated gluon density described here}
  \label{Fig:updf-uncertainty}
\end{figure}

The uncertainties of the pdf are obtained within the  \verb+herafitter+ package from a variation of the individual parameter uncertainties following the procedure described in \cite{Pumplin:2001ct} applying $\Delta \chi^2=1$. The uncertainties on the gluon are small (much smaller than obtained in standard fits), since only the gluon density is fitted. The uncertainty bands for the gluon density are shown in fig.~\ref{Fig:updf-uncertainty}(left).
In fig.~\ref{Fig:updf-uncertainty}(right) the prediction for $b$-jet cross section as calculated from  \CASCADE\ \protect\cite{Jung:2010si}  using the gluon density described here (labeled as {\bf set0}) is shown together with a prediction using an older set (labeled as {\bf setA 0}~\cite{Jung:2004gs})  in comparison with a measurement from CMS~\cite{Chatrchyan:2012dk}.

\vskip 0.5 cm 

\noindent 
{\bf Acknowledgments}.   We  thank  the conveners    for the invitation  and 
excellent organization of the  meeting.  

\begingroup\raggedright\endgroup



\begin{thebibliography}{10}

\bibitem{Aaron:2009aa}
F.~Aaron {\em et~al.}
\newblock \href{http://dx.doi.org/10.1007/JHEP01(2010)109}{JHEP {\bfseries
  1001} (2010) 109}, \href{http://arxiv.org/abs/0911.0884}{{\ttfamily
  arXiv:0911.0884 [hep-ex]}}.
61 pages, 21 figures.

\bibitem{Jung:2002wn}
H.~Jung.
\newblock Acta Phys. Polon. {\bfseries B33} (2002) 2995--3000,
\href{http://arxiv.org/abs/hep-ph/0207239}{{\ttfamily arXiv:hep-ph/0207239}}.

\bibitem{Hansson:2003xz}
M.~Hansson and H.~Jung.
\newblock
\href{http://arxiv.org/abs/hep-ph/0309009}{{\ttfamily arXiv:hep-ph/0309009}}.

\bibitem{Catani:1990eg}
S.~Catani, M.~Ciafaloni, and F.~Hautmann.
\newblock
\href{http://dx.doi.org/10.1016/0550-3213(91)90055-3}{Nucl. Phys. {\bfseries
  B366} (1991) 135}.

\bibitem{Jung:2010si}
H.~Jung, S.~Baranov, M.~Deak, A.~Grebenyuk, F.~Hautmann, {\em et~al.}
\newblock \href{http://dx.doi.org/10.1140/epjc/s10052-010-1507-z}{Eur.Phys.J.
  {\bfseries C70} (2010) 1237},
\href{http://arxiv.org/abs/1008.0152}{{\ttfamily arXiv:1008.0152 [hep-ph]}}.

\bibitem{Kwiecinski:1996td}
J.~Kwiecinski, A.~D. Martin, and P.~Sutton.
\newblock \href{http://dx.doi.org/10.1007/s002880050206}{Z.Phys. {\bfseries
  C71} (1996) 585}, \href{http://arxiv.org/abs/hep-ph/9602320}{{\ttfamily
  arXiv:hep-ph/9602320 [hep-ph]}}.

\bibitem{Ciafaloni:1987ur}
M.~Ciafaloni.
\newblock
\href{http://dx.doi.org/10.1016/0550-3213(88)90380-X}{Nucl. Phys. {\bfseries
  B296} (1988) 49}.

\bibitem{Andersson:1995ju}
B.~Andersson, G.~Gustafson, and J.~Samuelsson.
\newblock \href{http://dx.doi.org/10.1016/0550-3213(96)00114-9}{Nucl.Phys.
  {\bfseries B467} (1996) 443}.
Revised version.

\bibitem{Andersson:2002cf}
B.~Andersson {\em et~al.}
\newblock \href{http://dx.doi.org/10.1007/s10052-002-0998-7}{Eur. Phys. J.
  {\bfseries C25} (2002) 771},
\href{http://arxiv.org/abs/hep-ph/0204115}{{\ttfamily arXiv:hep-ph/0204115}}.

\bibitem{Marchesini:1990zy}
G.~Marchesini and B.~R. Webber.
\newblock
\href{http://dx.doi.org/10.1016/0550-3213(91)90389-F}{Nucl. Phys. {\bfseries
  B349} (1991) 617}.

\bibitem{Marchesini:1992jw}
G.~Marchesini and B.~R. Webber.
\newblock
\href{http://dx.doi.org/10.1016/0550-3213(92)90181-A}{Nucl. Phys. {\bfseries
  B386} (1992) 215}.

\bibitem{Aaron:2009kv}
F.~Aaron {\em et~al.}
\newblock Eur.Phys.J. {\bfseries C64} (2009) 561,
  \href{http://arxiv.org/abs/0904.3513}{{\ttfamily arXiv:0904.3513 [hep-ex]}}.
35 pages, 10 figures.

\bibitem{herafitter}
``\mbox{HERAFitter}'', 2012.
\newblock \url{http://herafitter.hepforge.org/}.

\bibitem{Deak:2010gk}
M.~Deak, F.~Hautmann, H.~Jung, and K.~Kutak.
\newblock
\href{http://arxiv.org/abs/1012.6037}{{\ttfamily arXiv:1012.6037 [hep-ph]}}.

\bibitem{Jung:2004gs}
H.~Jung.
\newblock
\href{http://arxiv.org/abs/hep-ph/0411287}{{\ttfamily arXiv:hep-ph/0411287}}.

\bibitem{Pumplin:2001ct}
J.~Pumplin, D.~Stump, R.~Brock, D.~Casey, J.~Huston, {\em et~al.}
\newblock \href{http://dx.doi.org/10.1103/PhysRevD.65.014013}{Phys.Rev.
  {\bfseries D65} (2001) 014013},
\href{http://arxiv.org/abs/hep-ph/0101032}{{\ttfamily arXiv:hep-ph/0101032
  [hep-ph]}}.

\bibitem{Chatrchyan:2012dk}
S.~Chatrchyan {\em et~al.}
\newblock JHEP {\bfseries 1204} (2012) 084,
\href{http://arxiv.org/abs/1202.4617}{{\ttfamily arXiv:1202.4617 [hep-ex]}}.

\end{thebibliography}
\end{document}